\documentclass[useAMS,usenatbib,]{mnras}
\usepackage{graphicx}
\usepackage{natbib}
\usepackage{bm}
\usepackage{amsmath}
\usepackage{amssymb}
\usepackage{mathtools} 
\usepackage{tabularx}

\newcommand{\oder}[2]{\frac{d #1}{d #2}}

\newcommand{\beq}{\begin{equation}}
\newcommand{\eeq}{\end{equation}}

\newcommand{\Mb}{\mathfrak{M}}

\title[Centrifugal instability of compressible flows]{Centrifugal instability of compressible flows and the hydrodynamic stability of accretion disks}
\author[S.S.Komissarov and K.N.Gourgouliatos]{
Serguei S. Komissarov,$^{1}$\thanks{E-mail: s.s.komissarov@leeds.ac.uk (SSK)}
and Konstantinos N. Gourgouliatos,$^{2}$\thanks{E-mail: kngourg@upatras.gr (KNG)}
\\
$^{1}$School of Mathematics, University of Leeds, LS2 9JT, UK\\
$^{2}$Laboratory of Universe Sciences, Department of Physics, University of Patras, Patras, Rio, 26504, Greece
}

\begin{document}
\date{Received/Accepted}
\maketitle
                                                                                                
\begin{abstract}

A recent analysis of the centrifugal instability in the case of pressure-supported compressible relativistic rotation, with application to astrophysical jets,  yielded a generalisation of the famous Rayleigh criterion for Newtonian flows. According to this criterion, the centrifugal instability is strongly affected by the flow Mach number, and  not only in the relativistic fluid dynamics but also in its Newtonian limit. To validate the Newtonian version of this criterion, we performed axisymmetric numerical simulations of non-relativistic transonic rotating flows which are stable according to the original Rayleigh criterion but can be either stable or unstable according to the new one. The results of computer simulations are found to be in perfect agreement with the theory.  The hydrodynamic stability of accretion disks is often explained by referring to the original Rayleigh criterion, even if their rotation is highly supersonic. To clarify the matter, we  analysed the hydrodynamic stability of  flows rotating about central compact object and derived an instability  criterion that  retains the explicit dependence on the flow Mach number.  This criterion turns out to be equivalent to the standard  Solberg-H{\o}iland criterion, which does not involve the Mach number. The same applies to the case of pressure-supported rotation, where the role of gravity is played by the centrifugal force. 

\end{abstract}   
                                                                                       

\section{Introduction}
\label{sec:intro}

The centrifugal instability was originally identified in studies of rotating flows where the centrifugal force is balanced by the pressure gradient. A perturbation of such equilibrium normally disturbs this balance resulting either in the restoring total force (stable configuration) or the total force driving the system further away from the equilibrium (unstable) configuration.  For inviscid incompressible fluid of constant density, the instability arises when the specific angular momentum decreases with the radius \cite{Rayleigh:1917}.   Later, it was found that the instability is not confined to the case or rotating fluid but may develop in other types of flows involving curved streamlines, such as flows past concave wall, and flows with curved free shear layers \citep[e.g.][]{Gortler55,Drazin:1981,Saric:1994,Otto94,Li2010,Boiko2010,Gourgouliatos:2018a,Sahli20}.  In all these cases, the instability develops when the flow is locally similar to rotation satisfying Rayleigh's instability criterion \citep[cf.][]{Bayly88}.  In the case of flows past concave walls, the favourable velocity profile is generated by the viscosity inside the boundary layer.  Typically, the growing perturbations can be described as steam-wise directed vortices giving rise to development of finger-like or mushroom-like structures in the non-linear regime, and facilitating transition to turbulence.      

The centrifugal instabilities are important for many industrial applications and have been extensively studied in the context of such applications.  They are known to occur in oceanic and atmospheric flows, playing an important role in their mixing and energy dissipation \citep[e.g.][]{GMM16,TSV18}.  In these applications, there is a multitude of factors influencing the stability, such as viscosity, thermo-conductivity, gravity, Coriolis force, buoyancy force, magnetic field etc.         

In astrophysics, the centrifugal force is often an important factor affecting the structure, dynamics and stability of stars and accretion disks \citep[e.g.][]{GS67,Fricke68,Seguin:1975,Tassoul-00,FKR-02,AF-13}.  
Recently, it emerged that the centrifugal instability can be important  in the dynamics of astrophysical jets as well. These jets are produced by compact objects,  stars or black holes, whose gravity dominates in their vicinity and promotes rapid pressure decline of the surrounding gas. While propagating through such gas the jets become free-expanding, lose causal connectivity and thus become very stable. This explains how they can propagate to such huge distances from their central engine \citep{PK15}.  However, they may eventually reach the distances where the gravity of their engine is no longer dominant and the external gas pressure distribution flattens out. There the pressure of external gas becomes dynamically important  and drives recollimation/reconfinement shock into the jets. In an alternative scenario, the jet may inflate an over-pressured cavity in the external gas confined by the ram-pressure of the bow shock it drives through the external gas.   In this scenario, it is the cavity pressure that is responsible for the jet recollimation.   In both the cases, downstream of the recollimation shock the jet is no longer rapidly expanding and becomes susceptible to various fluid instabilities.  \citet{PK15} argued that in the case of AGN jets it is the jet power that determines which of the scenarios applies.     \citet{Gourgouliatos:2018a} used 3D relativistic hydrodynamic simulations to explore the jet response to the recollimation and discovered that in both scenarios the recollimation is followed by a rapid transition to turbulence accompanied by efficient mixing between the jet and the external gas. Quite remarkably, in the transition between the laminar and turbulent sections of the jet, the numerical solution exhibited non-linear perturbations aligned with the jet streamlines, never seen in any of the previous jet simulations. Later this behaviour was reproduces in the simulations by other groups \citep{Gottlieb21,Costa26}.             

Since in the outer shocked layer of the jet its stream lines are curved,  \citet{Gourgouliatos:2018a} argued that the dominant instability of recollimating jets observed in these simulation is the centrifugal instability.  In the accompanying paper, they used a heuristic approach to derive the relativistic generalisation of Rayleigh's criterion for the centrifugal instability in the case of axisymmetric  rotating flows \citep{Gourgouliatos:2018b}. 

In the Newtonian limit, the criterion has a rather unusual form, 
\begin{equation}
    \frac{d \ln \Psi}{d \ln R}<M^2\,,
 \label{eq:GRC}  
\end{equation}
where $R$ is the cylindrical radius,  $\Psi = \rho \left(v_{\phi} R \right)^2$, where $\rho$ is the fluid density,  $v_\phi=\Omega R$ is the speed of rotation, $\Omega$ is its angular velocity, and $M=v_\phi/a$ is its Mach number. In the rest of the paper we refer to it as the KG criterion.  

In the limit $M\to0$ and constant $\rho$, \eqref{eq:GRC} reduces to the familiar Rayleigh criterion for incompressible fluid,
\begin{eqnarray}
\frac{d}{dR}\left(v_\phi R \right)^2<0\,.
\label{eq:Rayleigh}
\end{eqnarray}
Hence the only truly new feature of the centrifugal instability for compressible flows uncovered by the KG criterion  is its dependence on the flow Mach number. 

The dependence is quite strong and deserves special attention.  When the velocity of rotation $v_\phi\propto R^{-n}$ and $\rho$ is constant, the KG instability criterion yields 
\begin{equation}
    n>1-\frac{M^2}{2}\,.
 \label{eq:powerlaw}  
\end{equation}
Thus for $M>\sqrt{2}$, the equilibrium is unstable for any positive power index $n$, including $n=1/2$ of the Keplerian motion.  This is in conflict with the common claim that thin accretion disks, whose orbital velocity is hypersonic, are stable with respect to the centrifugal instability simply because the original Rayleigh criterion  \eqref{eq:Rayleigh} is not satisfied. 

In this paper, we use computer simulations of axisymmetric compressible rotating Newtonian flows to verify the dependence of the centrifugal instability on the flow Mach number.

\section{Method}
\label{sec:method}

In our simulations, we numerically solve the equations of ideal compressible fluid dynamics comprising of the continuity equation: 
\begin{eqnarray}
    \partial_t \rho +\nabla \cdot\left(\rho \bm{v}\right)=0\,,
\end{eqnarray}
the momentum equation:
\begin{eqnarray}
    \partial_t \left(\rho \bm{v}\right)+\nabla \cdot\left(\bm{v}\otimes \rho \bm{v}\right)+\nabla p=\bm{0}\,,
\end{eqnarray}
and the energy equation
\begin{eqnarray}
    \partial_t e + \nabla\cdot \bm{v}(\frac{1}{2}\rho v^2 + \rho h)=0\,,
\end{eqnarray}
where $h$ is the specific enthalpy. These equations are complemented with the equation of state of ideal gas  $p = K \rho ^{\gamma}$. The corresponding specific enthalpy 
\begin{eqnarray}
    h=\frac{\gamma}{\gamma-1}\frac{p}{\rho}\,,
\end{eqnarray}
and the sound speed 
\begin{eqnarray}
    a=\sqrt{\gamma\frac{p}{\rho}}\,.
    \label{eq:sound_speed}
\end{eqnarray}
In the simulations, the ratio of specific heats $\gamma=5/3$.

The numerical simulations were performed using the HD version of the AMR-VAC code \citep{Keppens:2012,Porth:2014}.  The key built-in routines for integration included HLL, Koren flux limiter and 4th-order RK time integrator.  For selected models, the convergence of numerical solutions was checked via doubling the resolution.

\section{Simulation setup}
\label{sec:Simulation Setup}

In the numerical simulations, we constrain the problem to the case of axisymmetric flows. The initial equilibrium solution depends only on the radial coordinate, and hence the force balance dictates    
\begin{eqnarray}
\frac{dp}{dR}=\frac{\rho v_{\phi}^2}{R}\,.
\label{eq:equilibrium}
\end{eqnarray}
For simplicity,  we also assume uniform density distribution $\rho(R)=\rho_0$. In this case, equation (\ref{eq:equilibrium}) yields  the pressure distribution
\begin{eqnarray}
    p(R) = p_0 +\rho_0\int_{R_0}^R  \frac{v_{\phi}^2(\tilde{R})}{\tilde{R}}d{\tilde R}\,,
\end{eqnarray}
where $p_0=p(R_0)$.
Here we adopt the configuration with $R_0=0$, the solid body rotation in the central core of radius $R_c$, and the power law $v_\phi(R)\propto R^{-n}$ with $n>0$ outside of the core,  namely 
\begin{eqnarray}
    v_{\phi}=\begin{dcases} \Omega_0 R\,, & 0\leq R\leq R_c \\
    \Omega_0 R_c \left(\frac{R}{R_c}\right)^{-n}\,, & R>R_c\,,
\end{dcases}
\label{eq:init-sol}
\end{eqnarray}
where $\Omega_0$ is the angular velocity of the core.
The corresponding pressure distribution is 
\begin{eqnarray}
    p=\begin{dcases} p_0+\frac{\rho_0 \Omega_0^2 R^2}{2}\,,  &   R\leq R_c \\
    p_0+\frac{\rho_0\Omega_0^2 R_c^2}{2} f(R/R_c)\,,  &     R>R_c\,,
\end{dcases}
\end{eqnarray}
where 
$$
f(x)= 1 +\dfrac{1}{n}\left(1 - x^{-2n}\right)\,.
$$ 
It is easy to see that the Mach number of the flow has a local maximum at $R=R_c$, where it reaches the value
\begin{eqnarray}
    M_c= \left( \frac{\gamma p_0}{\rho_0\Omega_0^2R_c^2} + \frac{\gamma}{2}\right)^{-1/2}\,.
\end{eqnarray}
It is immediately clear that $M_c \leq \sqrt{2/\gamma}$, and thus this configuration allows us to explore at most transonic flows. 
To overcome this constraint, we also explored the setup with $0<R_0<R_c$, but because the results of that study did not reveal anything particularly different they are not presented here.    

Inside the core,
\begin{eqnarray}
    \frac{d\ln \Psi}{d\ln R} = 4,
\end{eqnarray}
and hence the KG criterion predicts stability of its flow. 

\begin{figure*}
\begin{center}
     \includegraphics[width=0.45\textwidth]{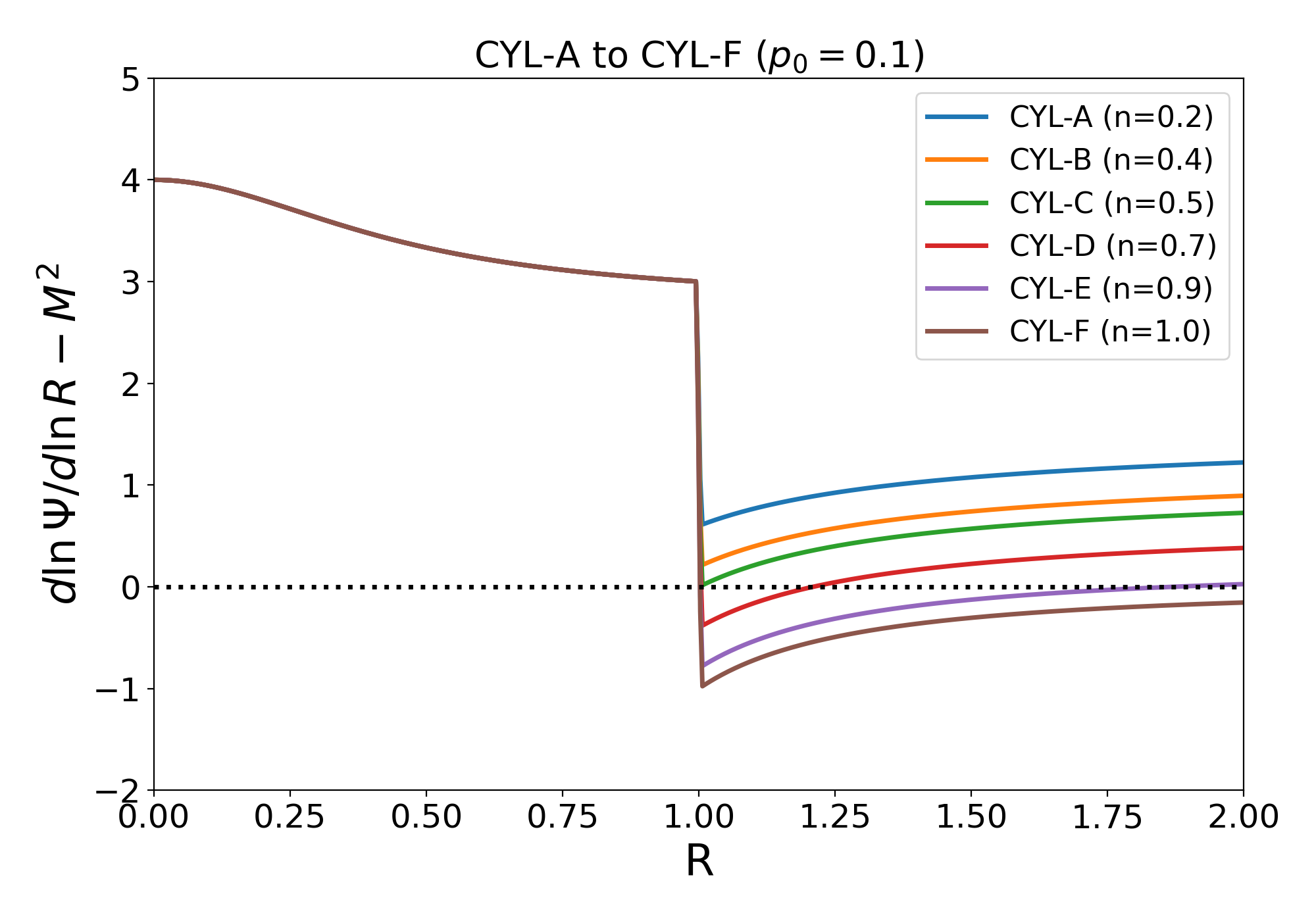}
     \includegraphics[width=0.45\textwidth]{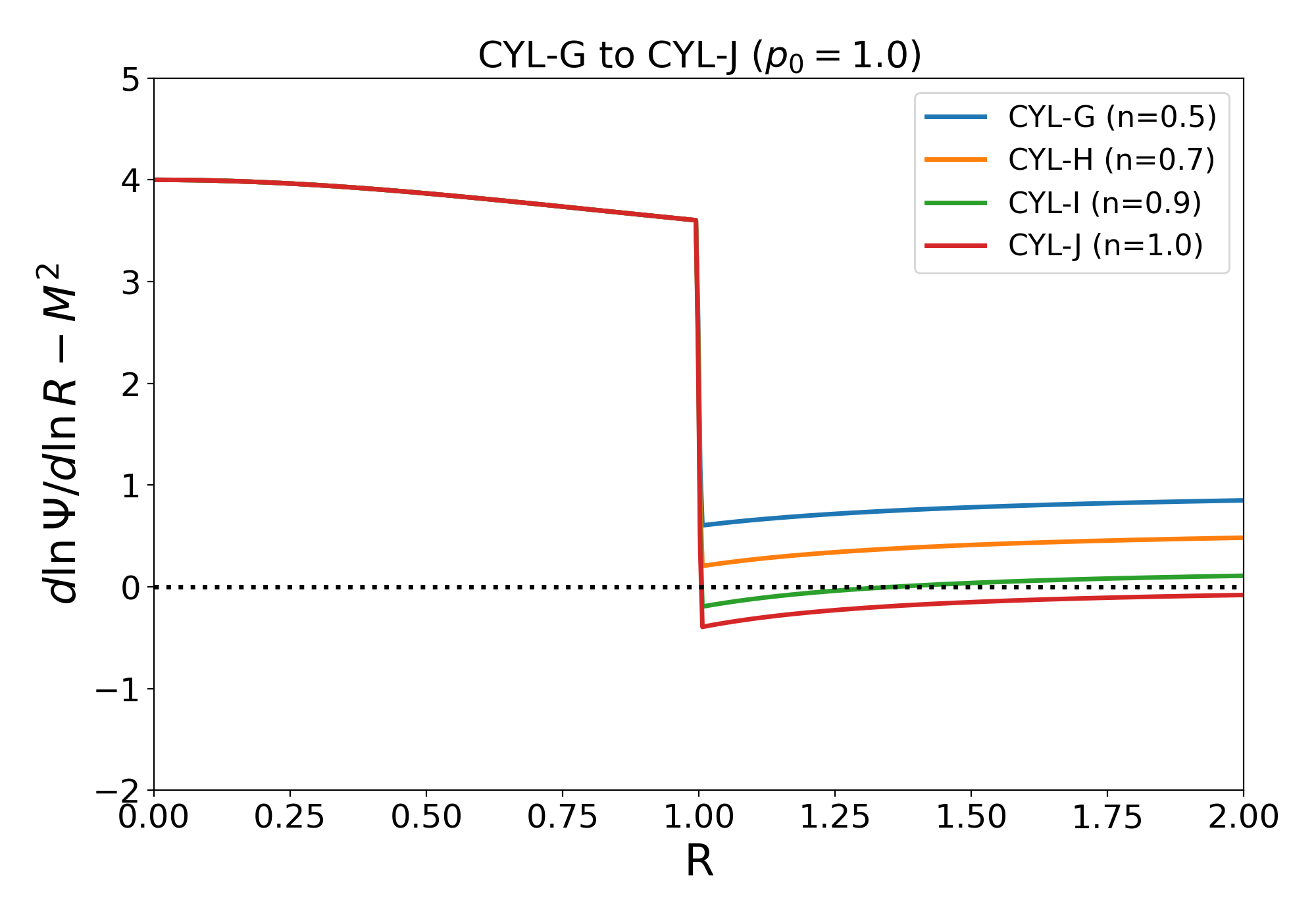}     
\end{center}
    \caption{Stable and unstable domains in the models with $p_0=0.1$ (top panel) and in the models with $p_0=1.0$ (bottom panel). In the unstable domains, $d\ln\Psi/d\ln R-M^2<0$.} 
    \label{fig:1}
\end{figure*}

Outside of the core, $d\ln\Psi/d\ln{R} =2(1-n)$, and the instability criterion reduces to 
\begin{eqnarray}
    2(1-n)<M^2\,.
    \label{eq:criterion}
\end{eqnarray} 
According to this equation, the flow is expected to be unstable for all $R>R_c$ provided $n\ge1$. For $n<1$, the flow may be either unstable in a vicinity of the core and stable further out, or stable everywhere outside of the core. Because for $R>R_c$ the flow Mach number monotonically approaches zero, this is decided its value at $R=R_c$. If $M_c$ exceeds $2(1-n)$, the flow will be unstable inside some finite interval $[R_c,R_c\Delta R]$ and stable further out.  If $M_c$ is below $2(1-n)$, the flow will be stable for all $R>R_c$. If  $M_c=2(1-n)$, we have a marginally stable configuration.   Figure \ref{fig:1} shows the distributions of $d\ln\Psi/d\ln R - M^2$ for the parameters used in the simulations, namely $\rho_0=1$, $\Omega_0=1$, $R_c=1$, $p_0=0.1$ and $1.0$, and the power index $n$ ranging from 0.2 to 1.0.  Table \ref{tab:1} lists the simulations models, gives their key parameters and the expected domains of instability.

\begin{table}
\begin{center}
\caption{The key parameters of the simulation models. In the last column, ''S'' means that the flow was found stable in the simulations and ''U'' that it was found unstable.  
 In all the models, $\rho_0=1$,  $\Omega_0=1$, and $R_c=1$.}
\label{tab:1}
\begin{tabular}{|c|c|c|c|c|c|}
 \hline
 &&&&expected&\\
 model & $n$ & $M_c$  & $p_0$ & instability & stability\\ 
 & & & & domain & indicator \\
 \hline
 &&&&&\\
CYL-A & 0.2  & 1    & 0.1  & $-$ & S\\

CYL-B & 0.4  & 1     & 0.1  & $-$ & S\\

CYL-C & 0.5  & 1   & 0.1 & $[1,1]$ & U  \\

CYL-D & 0.7  & 1  & 0.1    & $[1,1.17]$ & U\\

CYL-E &0.9  & 1    & 0.1  & $[1,1.78]$ & U\\

CYL-F & 1.0  & 1   & 0.1  & $[1,2.00]$ & U\\

CYL-G & 0.5  & 0.63  & 1.0 & $-$ & S \\

CYL-H & 0.7  & 0.63 &  1.0  & $-$ & S\\

CYL-I & 0.9  & 0.63     & 1.0  & $[1,1.32]$ & U\\

CYL-J & 1.0  & 0.63   & 1.0  & $[1,2.00]$ & U \\
&&&&&\\
\hline
\end{tabular}
\end{center}
\end{table}

The original Rayleigh instability criterion is also not satisfied in the core, whereas outside of the core it reduces to 
\beq
    2(1-n)<0\,.
\eeq   
Hence, for all values of $n$ used in the simulations, with the possible exception for $n=1$, the Rayleigh criterion predict stability of the flow. For $n=1$, it predicts a marginally stable flow.   

The simulation domain is $(r,z)\in[0,2]\times[0,1]$, with the base resolution of $2000\times1000$ cells. The periodic boundary conditions are imposed at $z=0$ and $z=1$. The reflective boundary conditions are imposed at $R=0$ and $R=2$. The equilibrium is perturbed via the velocity perturbation 
\begin{eqnarray}
   \delta v_{\phi} =\sin\left(k z\right) 
   \label{eq:pert1}
\end{eqnarray}
with the amplitude $\delta v_0=10^{-4}\Omega_0 R_c$ and the wavenumber $k=10\pi$ ($\lambda=0.2$).

\section{Results}
\label{sec:Results}

The main aim of these simulations was to investigate whether the onset of the centrifugal instability occurs in agreement with the KG criterion, which for the used initial equilibrium state \eqref{eq:init-sol} reduces to equation \eqref{eq:criterion}.  In the last column of table \ref{tab:1}, symbol "U" means the instability was detected in the simulations, whereas the symbol "S" means that it was not. Comparing the outcome with the theoretical prediction, based on the instability criterion for compressible flow and given in the penultimate column of this table, shows full agreement between the two.        

\begin{figure*}
     \includegraphics[width=0.32\textwidth]{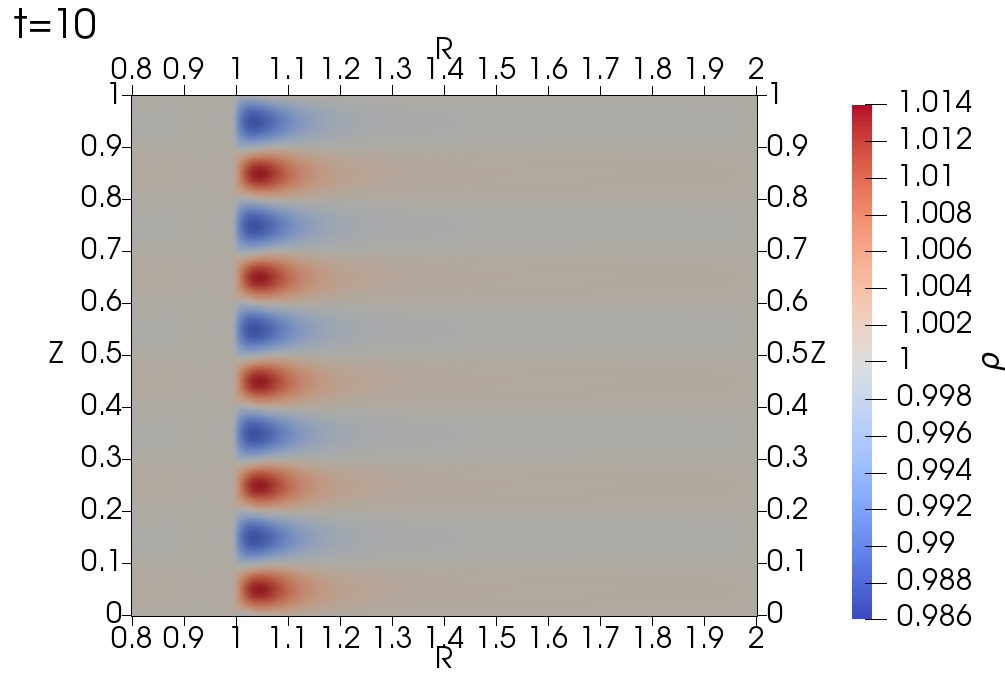}
     \includegraphics[width=0.32\textwidth]{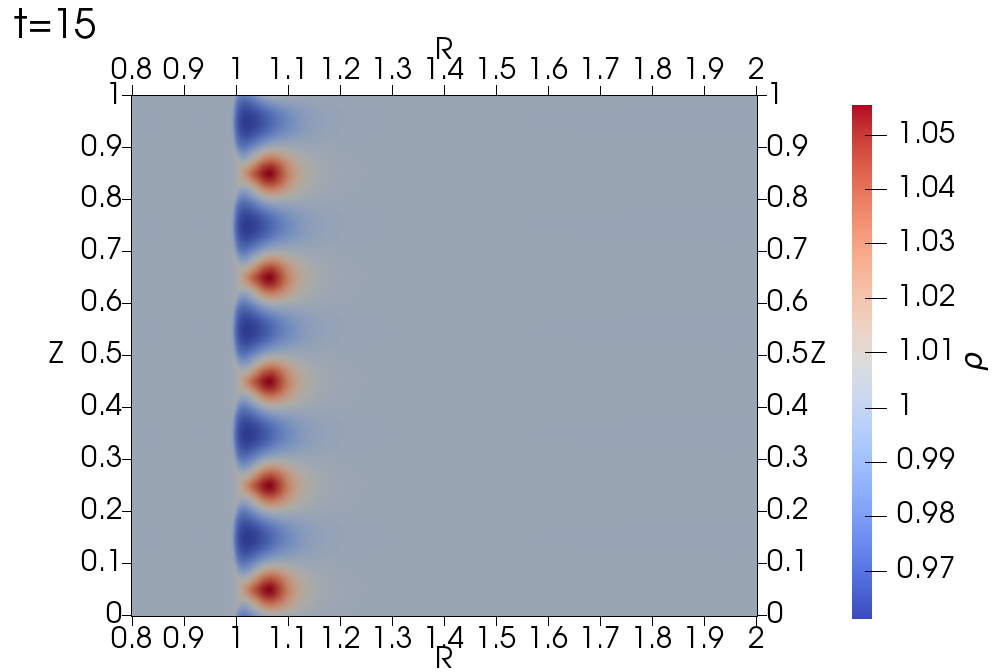} 
     \includegraphics[width=0.32\textwidth]{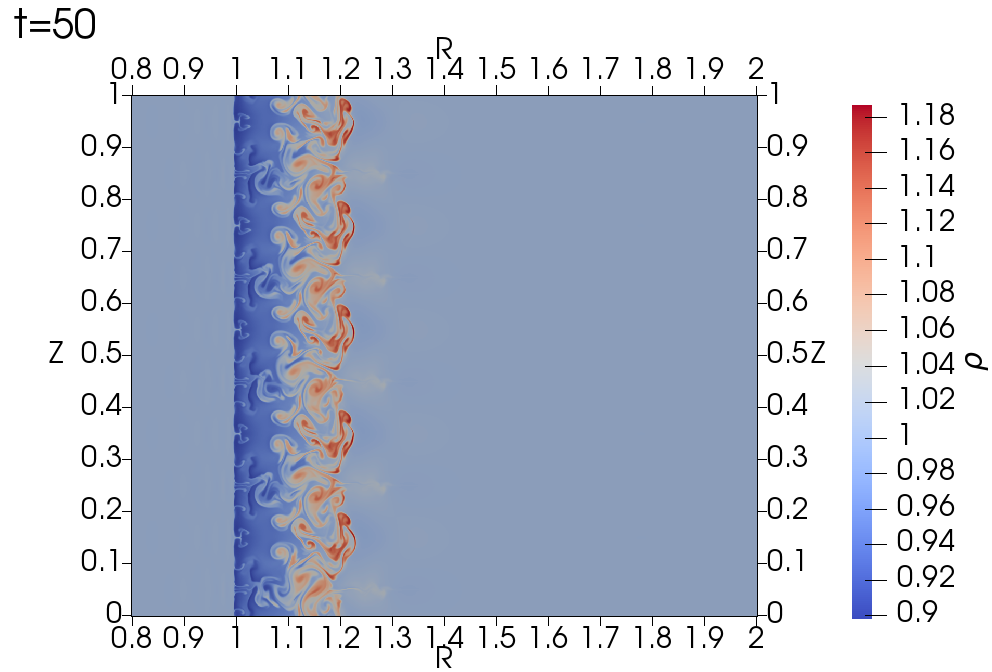}
     \includegraphics[width=0.32\textwidth]{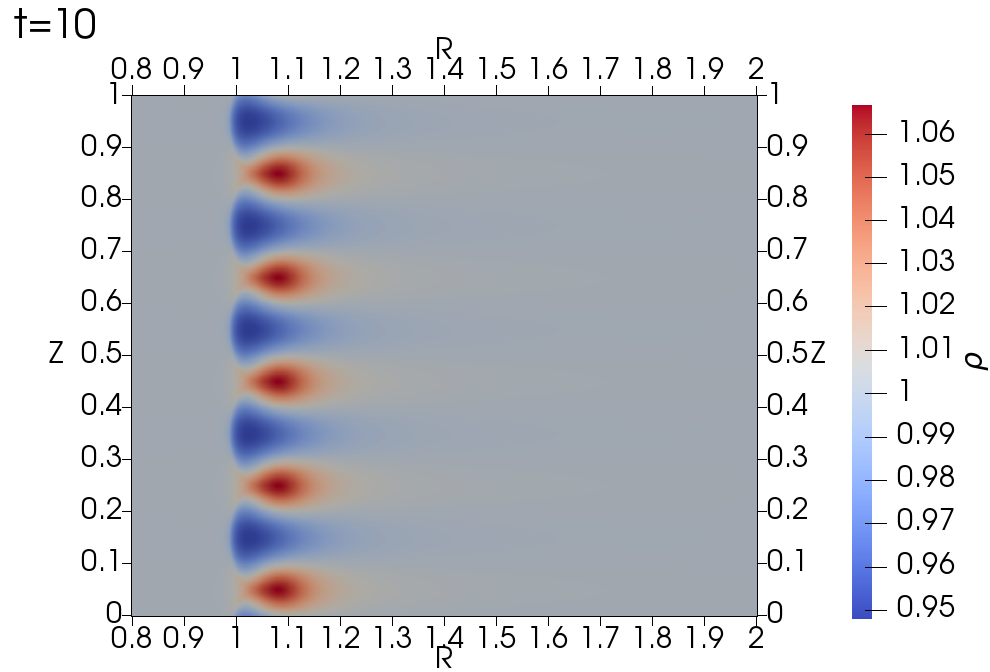}
     \includegraphics[width=0.32\textwidth]{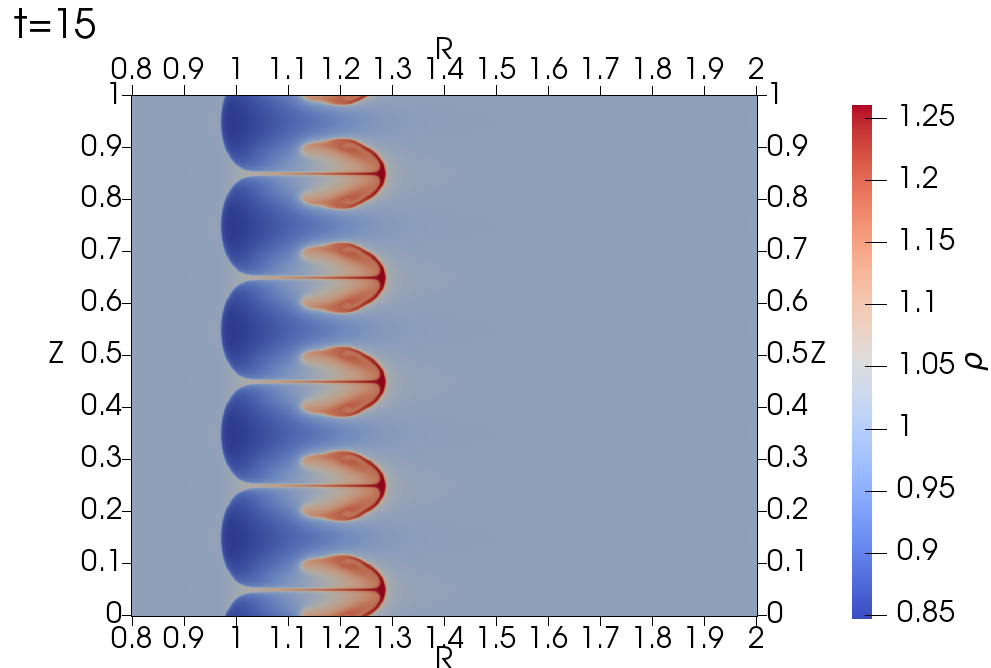} 
     \includegraphics[width=0.32\textwidth]{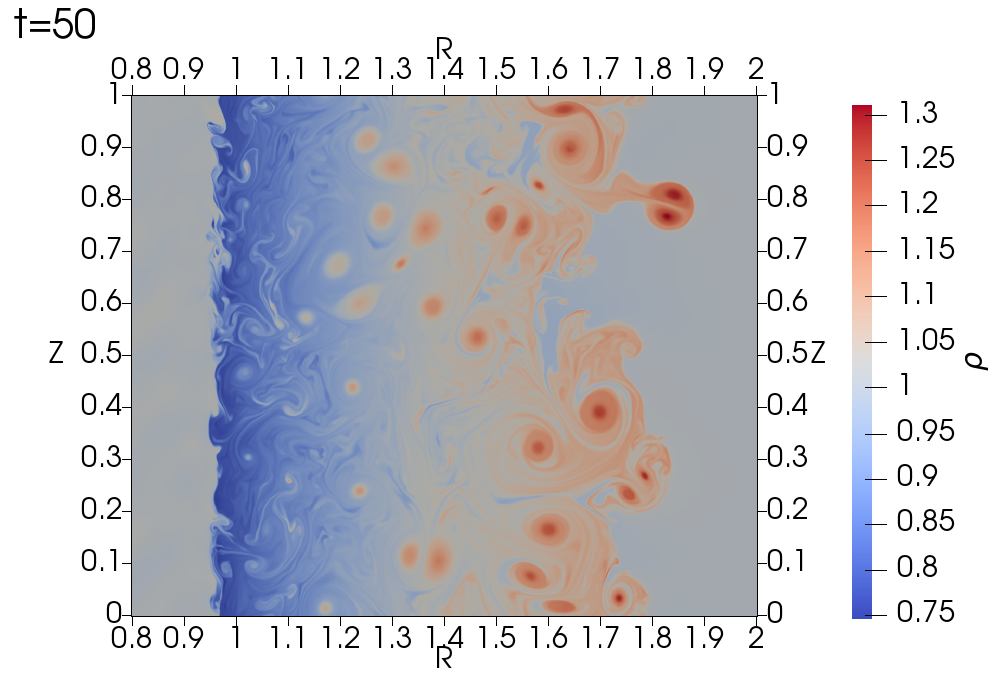}
     \includegraphics[width=0.32\textwidth]{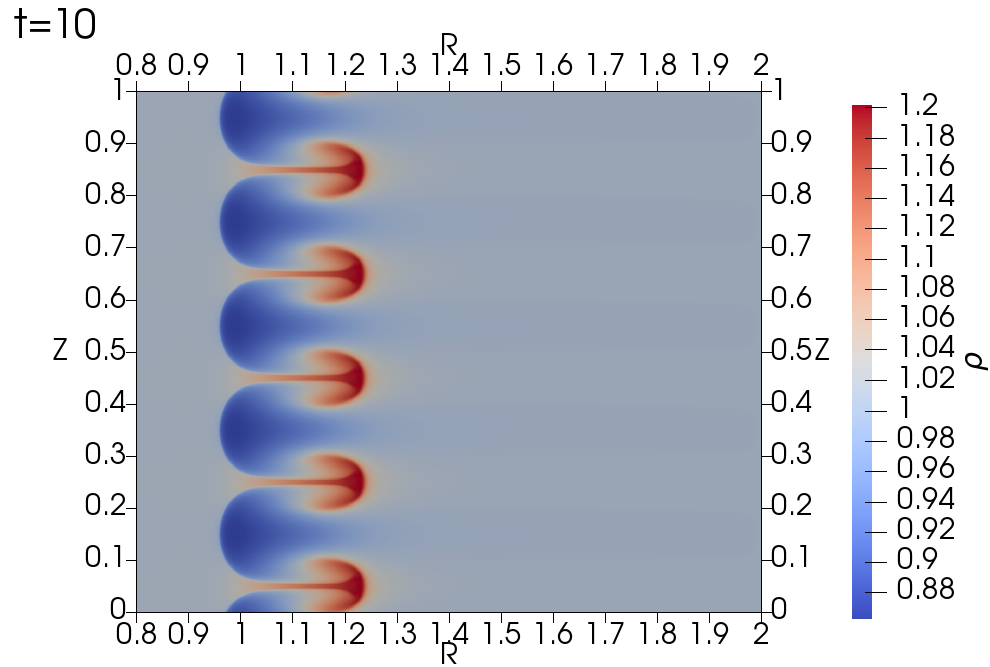}
     \includegraphics[width=0.32\textwidth]{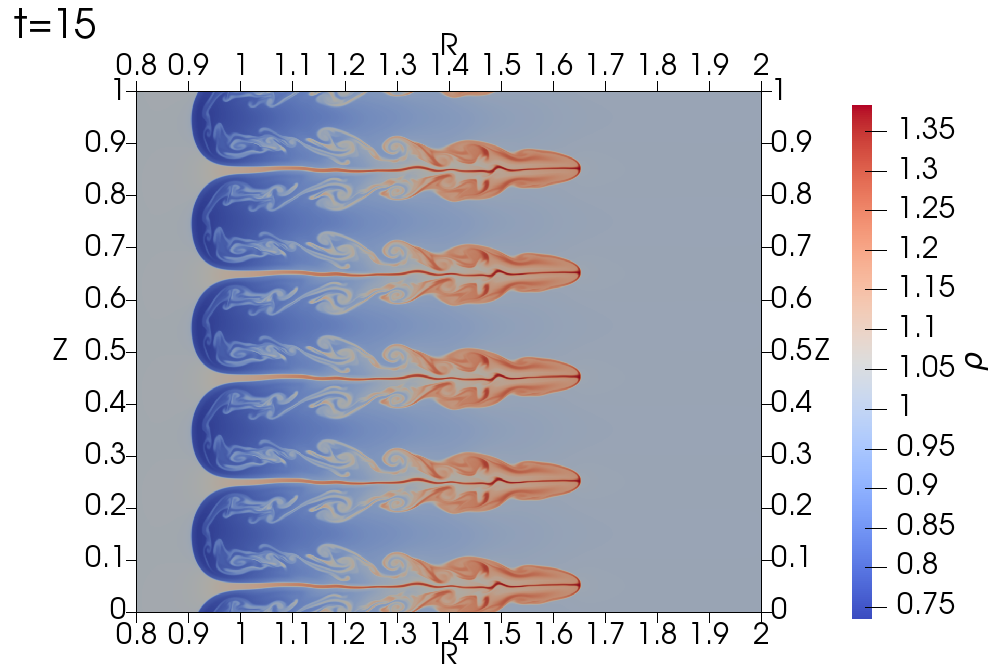} 
     \includegraphics[width=0.32\textwidth]{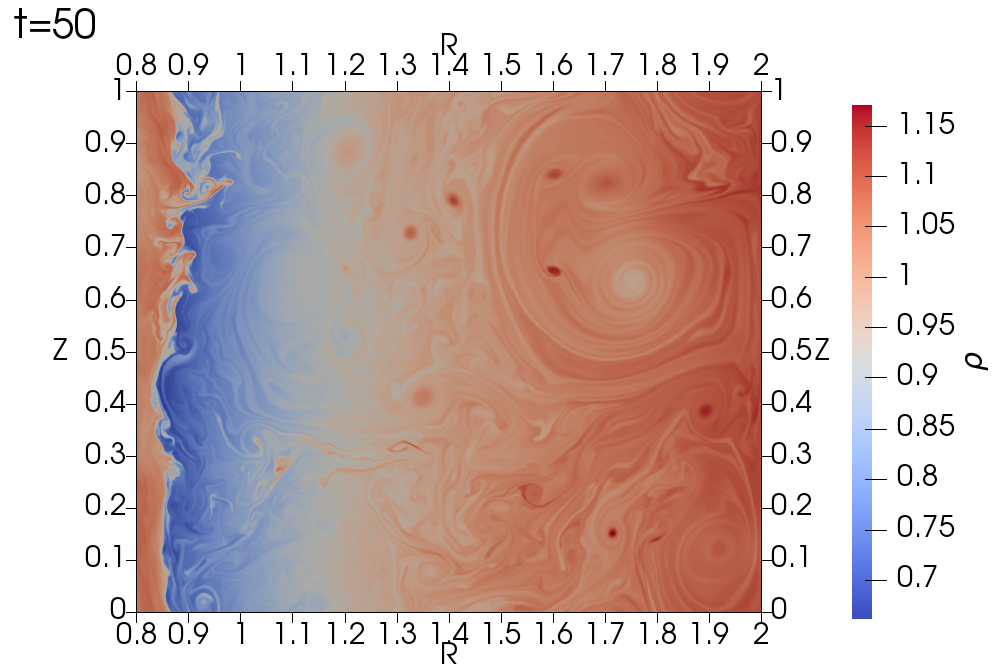}
 \caption{The density distribution at $t=10,~15,~50$ for the model CYL-C (top panels),  CYL-D (middle panels) and CYL-F (bottom panels). } 
    \label{fig:2}
\end{figure*}

A proper study of the nonlinear phase of the instability was not intended as this requires full 3D simulations.  So here we only outline our naked-eye observations.  The instability apparently grows at fastest rate at inner boundary of the instability domain ($R=1$ in all models),  where $2(1-n)-M^2$ has its lowest value.  In figure \ref{fig:2}, this is evident in the solutions for the models CYL-C and CYL-D at $t=10$, where the perturbation is still quite weak. 

As the perturbation grows to the fully nonlinear amplitude,  the flow develops streams (fingers) moving radially away from the core. These streams interact with the surrounding fluid and create the fungi-like structures seen in figure \ref{fig:2}  at $t=15$ for the model CYL-D, and at $t=10$ for the model CYL-F.  Later, the fungi hats turn into cocoons surrounding the outflowing streams and exhibiting the rolls characteristic to the Kelvin-Helmholtz instability. These structures are reminiscent of numerical fluid jets studied in connection to the jets of active galactic nuclei. The jets may extend well beyond the instability domain and create an extended zone of turbulent flow. For example, the instability domain of the model CYL-D is $[1,1.17]$, whereas the turbulent zone extends up to $R=2$ (see figure \ref{fig:2}).          

%

\section{Stability of accretion disks} 
\label{sec:disks}

Given the strong destabilising role of high Mach number in the centrifugal instability, it is only natural to enquire why this does not seem to matter in the stability of accretion disks where the rotational motion is highly supersonic.  Their hydrodynamic stability was a hot topic in astrophysics in connection to the origin of turbulence required for the angular momentum transport, which is needed to sustain the accretion flow. Now it is widely accepted that thin Keplerian accretion discs are hydrodynamically stable and only the inclusion of magnetic field leads to their instability \citep[e.g.][]{Chandra-60,Balbus:1991,BHS-96,Balbus-03}.

Quite often a short explanation of the stability is provided with the reference to the Rayleigh instability criterion \eqref{eq:Rayleigh}. Namely, because in Keplerian (thin) accretion disks $v_\phi\propto R^{-1/2}$, the criterion is not satisfied, and hence the disks are stable. Since in most textbooks on fluid instabilities \citep[e.g][]{Chandra61}, the centrifugal instability is discussed in the context of pressure-supported rotation of incompressible flows,  this explanation appears somewhat puzzling. Indeed, not only the fluid of  Keplerian accretion disks is compressible, and its rotation is supersonic, the centripetal force is not the gas pressure but the gravity of the accreting object. The pressure force acts in the same direction as the centrifugal force, and is quite weak. 

When this small pressure force is completely ignored, the stability problem becomes trivial.   Each fluid parcel of the disk moves like a free particle in the gravitational field of the central object,  and so the disk stability is essentially the same as the stability of circular orbits in Newtonian gravity. In particular, a radially displaced ring of the disk oscillates about its initial position with the epicyclic frequency $\Omega_e=\sqrt{G\Mb/R^3}$.  To elucidate the case with non-vanishing pressure force, which can be strong in thick accretion disks and rotating stars, one can  apply the same heuristic approach as in \citet{Landau87} and \citet{Gourgouliatos:2018b}.  While it does not involve solving a dispersion equation and hence does not yield any information on the instability growth rate and spectrum, it provides useful insights concerning the physical nature of instabilities.

In the equatorial plane of the disk (star), the equilibrium is described by the equation 

\beq
-\oder{p}{R} +\frac{\rho v_\phi^2}{R} - \rho\frac{G\Mb}{R^2} =0\,,
\label{eq:equilib}
\eeq       
which states the balance of the pressure, centrifugal and the gravity forces respectively.  Suppose a thin circular ring of the fluid is adiabatically displaced from $R=R_1$ to $R_2=R_1+dR$.  The gas pressure of the ring rapidly adjust to that in its surrounding and the total radial force acting on the ring becomes 

\beq
df_R=-\frac{1}{R_2}(\rho_2 v_{2,\phi} -\tilde{\rho}_1 \tilde{v}_{1,\phi} )+\frac{G\Mb}{R_2^2}(\rho_2-\tilde{\rho}_1)\,,
\label{eq:force}
\eeq      
where for any parameter $A$
\begin{align}
\nonumber
A_1&=A(R_1)\,,\\
\nonumber
 A_2&=A(R_2)=A_1+\oder{A}{R}(R_1)dR\,
\end{align}
are its values in the disk, and 
\begin{align}
\nonumber
\tilde{A}_1=A(R_1)+\tilde{dA}
\end{align}
is its value in the displaced ring. For adiabatic displacement, 
\beq
\tilde{d\rho} = \frac{\rho(R_1)}{\gamma p(R_1)} dp \,,
\label{eq:adiab}
\eeq
and the conservation of the angular momentum yields
\beq
\tilde{dv}_\phi = -\frac{v_\phi(R_1)}{R_1} dR\,.
\label{eq:am-cons}
\eeq
Substituting \eqref{eq:adiab} and \eqref{eq:am-cons} into \eqref{eq:force} yields
\beq
\oder{f_R}{R} =-\frac{\rho v_\phi^2}{R^2} \oder{\ln\Psi}{\ln{R}} + \frac{M^2}{R}\oder{p}{R}  
-\frac{G\Mb}{R^2} \frac{\rho}{\gamma}\oder{S}{R} \,.
\eeq 
For instability, the force derivative has to be positive, and hence the instability criterion reads 
\beq
\frac{\rho v_\phi^2}{R^2} \oder{\ln\Psi}{\ln{R}} - \frac{M^2}{R}\oder{p}{R}  
+\frac{G\Mb}{R^2} \frac{\rho}{\gamma}\oder{S}{R} <0 \,. 
\label{eq:dinst0}
\eeq
To understand the physical nature of the instability, it is instructive to consider the limits of vanishing gravity and vanishing rotation. 

In the limit of vanishing gravity, $\Mb\to 0$, we have the Rayleigh problem of pressure-supported rotation.  The equilibrium \eqref{eq:equilib} reads
\beq
-\oder{p}{R} +\frac{\rho v_\phi^2}{R} =0\,,
\eeq  
and the instability criterion \eqref{eq:dinst0} reduces to \eqref{eq:criterion}, as expected.

In the limit of vanishing rotation, $v_\phi\to 0$, the equilibrium \eqref{eq:equilib} reduces to the equation of hydrostatic equilibrium 
\beq
-\oder{p}{R} - \rho\frac{G\Mb}{R^2} =0\,.
\eeq  
Using this equation, the instability criterion \eqref{eq:dinst0} can be written as
\beq
\oder{S}{R} <0\,, 
\eeq
where $S=\ln(p\rho^{-\gamma})$ is the specific entropy of the fluid. This is the Schwarzschild criterion for convective instability written in terms of the entropy.  Thus the criterion \eqref{eq:dinst0} describes the competition between the centrifugal and convective instabilities.

In general, the equilibrium \eqref{eq:equilib} allows to write \eqref{eq:dinst0} as 
\beq
 \frac{\rho}{R^3} \oder{(v_\phi R)^2}{R} -\frac{1}{\gamma}\oder{p}{R}\oder{S}{R} < 0\,.
\label{eq:dinst}
\eeq 
This result can be found in \citet{Stewart-75}, \citet{Rudiger-02}, for example, and apparently in much earlier works by \citet{Solberg36} and \citet{Hoiland-41}.   For isentropic flows where $p=K\rho^{\gamma}$, equation \eqref{eq:dinst} reduces to the Rayleigh criterion \eqref{eq:Rayleigh}, which has a simple explanation.  Indeed, \citet{Rayleigh:1917} did not limit his analysis to incompressible pressure-supported rotational flows, but also considered a more general case of fluid with a barotropic equation of state, $\rho=\rho(p)$, in the presence of external potential force (e.g. gravity), obtaining the same instability condition \eqref{eq:Rayleigh} for both these cases. Obviously, the isentropic fluid is a particular case of barotropic fluids.

For thin disks with almost Keplerian rotation, $v_\phi\propto R^{-1/2}$, the second term in \eqref{eq:dinst} is negative and hence cannot drive instability.  In the $\alpha$-models of accretion disks \citep{SS73}, $S\propto R^{1/5}$ when the opacity is dominated by electron scattering, and $S\propto R^{1/2}$ when it is dominated by free-free absorption. Hence the first term in \eqref{eq:dinst} is also negative, and therefore these disks are stable to radial axisymmetric perturbations.   

For pressure-supported flows, equation \eqref{eq:dinst} can be written as 
\beq
 \frac{\rho}{R^3} \oder{(v_\phi R)^2}{R} -\frac{1}{\gamma}\frac{\rho v_\phi^2}{R}\oder{S}{R} < 0\,, 
\label{eq:dinst1}
\eeq 
showing that in the convective term the role of gravity is played by the centrifugal force. Ultimately, the KG criterion is a formulation of the Solberg-H{\o}iland criterion in terms of parameters arguably more suitable (traditional) for pressure-supported rotational flows and jets. 

Obviously, in this paper we considered only the simple case of ideal fluid, which is relevant to astrophysical jets and accretion disks.   
Readers interested in the role of viscosity, thermo-conductivity and radiation transport can consult with  \citet{GS67,Fricke68,APS93,RPL88,UB98,AU04} among others.

\section{Conclusions}
\label{sec:Conclusions}

In this paper, we presented a numerical study of the centrifugal instability for ideal compressible non-relativistic rotational flows, where the rotation is supported solely by the gas pressure. It was focused on the configurations that are stable according to the classical Rayleigh criterion but unstable according to the modified criterion for compressible flows derived in \citet{Gourgouliatos:2018b}. Using axisymmetric hydrodynamic simulations, we have explicitly tested the modified criterion and examined the nonlinear evolution of unstable flows. 

The results of the simulations reveal the dependence of the instability on the flow Mach number, which is in excellent agreement with the theoretical expectations.  At the nonlinear phase of the instability, the unstable regions of the flow develop radial streams of fluid reminiscent of the fingers of the Rayleigh-Taylor instability. These streams acquire   mushroom shape and then produce cocoons filled with rolls characteristic to the Kelvin-Helmholtz instability. Ultimately, they produce a turbulent layer with strong mixing.  The size of this layer may exceed significantly the size of the initial unstable region of the flow.  Although these axisymmetric simulations are quite sufficient to verify the destabilising role of Mach number, full 3D simulations are required to establish the properties of the nonlinear mixing triggered by the centrifugal instability.      

Although in this study we investigated only a rather artificial problem of pure rotational flow, the results are relevant to any problem where a high-speed flow develops curved stream lines supported by the pressure force. In astrophysics, such configurations are expected to emerge via interaction of supersonic winds and jets from compact objects with the surrounding plasma \citep[e.g.][]{Gourgouliatos:2018a}.  They are also relevant to high-speed aerodynamics, though the flow viscosity must also be taken in consideration in such problems. 

Supersonic rotation is typical for accretion disks of compact objects (stars and black holes). In the case of thin (Keplerian) disks the rotation is actually in the hyper-supersonic regime. To investigate the role of compressibility in the stability of accretion disks, we extended our heuristic stability analysis to the case where both the pressure force and the gravity of the central object is taken into account.  In the limit of vanishing gravity, the new instability criterion reduces to the criterion for the centrifugal instability of compressible flows, including its explicit dependence on the Mach number. In the limit of vanishing rotation, it reduces to the Schwarzschild criterion for convective instability.   Thus it describes the competition between the centrifugal and convective instabilities.     In general, the criterion can be written in the same form as the Solberg-H{\o}iland criterium, where the dependence of the flow Mach number is no longer explicit.  Only in the case of isentropic flow it reduces to the original Rayleigh criterion for the centrifugal instability. This result is consistent with \citet{Rayleigh:1917} who analysed flows with barotropic equation of state.

\section*{Acknowledgments}
The authors are grateful to the anonymous referee for their insightful comments which identified inconsistencies of the original manuscript.

  This work was supported by computational time granted by the National Infrastructures for Research and Technology S.A. (GRNET S.A.) in the National HPC facility - ARIS - under project ID
pr017008/simnstar2.   

\section*{Data Availability}     

The data underlying this article will be shared on reasonable request to the corresponding author.

\bibliographystyle{mnras}
\bibliography{bibliography}

\end{document}